\documentclass[aps,prb,preprint,superscriptaddress,showpacs,amsmath,amsfonts]{revtex4-1}

\usepackage{graphicx}

\begin{document}


\title{Ferromagnetic instability of interlayer floating electrons in quasi-two-dimensional electride Y$_2$C}


\author{Takeshi Inoshita}
\affiliation{Materials and Structures Laboratory, Tokyo Institute of Technology, Nagatsuta, Kanagawa 226-8503, Japan}
\affiliation{National Institute for Materials Science, Tsukuba, Ibaraki 305-0044, Japan}
\author{Noriaki Hamada}
\affiliation{Faculty of Science and Technology, Tokyo University of Science, Noda, Chiba 278-8510, Japan}
\author{Hideo Hosono}
\affiliation{Materials and Structures Laboratory, Tokyo Institute of Technology, Nagatsuta, Kanagawa 226-8503, Japan}



\date{\today}

\begin{abstract}
Ab initio electronic structure calculations show that the recently identified quasi-two-dimensional electride Y$_2$C is a weak itinerant ferromagnet or at least close to a ferromagnetic instability.   The ferromagnetism is induced by the electride electrons, which are loosely bound around interstitial sites and overlap with each other to form two-dimensional interlayer conduction bands.  The semimetallicity and two-dimensionality of the band structure are the key to understanding this ferromagnetic instability.
\end{abstract}


\maketitle

Introduction:  Electrides are unconventional ionic materials in which there is an intrinsic excess of valence electrons from the formal valence viewpoint. \cite{Dy1, Dy2, Pi,Pic,Ki,Mat}  The excess electrons {\it float} in the void space between ions and, if their energy band crosses the Fermi level, dominate the transport and magnetic properties.   Electrides can be conveniently classified by the dimensionality of the electronic structure of the floating electrons.  In zero-dimensional (0D) electrides, the floating electrons are localized at interstitial sites and form a narrow band, whereas 1D and 2D electrides have electrons floating inside filamentary channels and interlayer gaps, respectively.   For decades, all the known electrides were either 0D or 1D, but recently Lee \textit{et al.} \cite{Le} have demonstrated, through transport measurements and electronic structure calculations, that the layered dicalcium nitride Ca$_2$N is a 2D electride having a single conduction band.  The electrons in this band are well confined in the void space between Ca layers while being able to move freely along the layers.  The concentration of the 2D electrons was found to be one electron per unit cell (Ca$_2$N), in agreement with that deduced from the formal valence [Ca$_2$N]$^{+}\cdot$ e$^{-}$.  

This finding inspired a search for further 2D electrides using  ab initio calculations.  Walsh and Scanlon found Sr$_2$N and Ba$_2$N to be 2D electrides.\cite{Wa}   We carried out an extensive database screening followed by ab initio calculations to identify, in addition to Sr$_2$N and Ba$_2$N,  six carbides (Y$_2$C and 4$f$ lanthanide carbides Ln$_2$C where Ln=Gd, Tb, Dy, Ho, Er)  to be 2D electrides.\cite{In}   Tada \textit{et al.} \cite{Tad} investigated, in addition to  Sr$_2$N, Ba$_2$N and Y$_2$C, some 2D electrides not yet registered in current inorganic material databases.  Despite their identical crystal structure (anti-CdCl$_2$ structure with unit cell M$_2$X, space group $R \bar{3} m$), the nitrides M$_2$N and carbides M$_2$C significantly differ in their electronic structures.  Their magnetic properties are expected to differ accordingly. The nitrides have a single conduction band with the character of a 2D interlayer band and are thought to be  Pauli paramagnets.  The carbides are more complex and have two conduction bands.  The lanthanide carbides Ln$_2$C are ferromagnets with a weakly ferrimagnetic spin order.  Y$_2$C is a marginal and less clear-cut case requiring a careful investigation to determine its ground state.  A magnetization measurement found no ferromagnetic transition down to a temperature of 2~K, \cite{Zh} which led us to report only spin-nonpolarized calculations for Y$_2$C in our previous paper. \cite{In}  

In the present paper, we investigate the electronic structure of Y$_2$C in detail and show that its ground state, as determined by the density functional and hybrid functional methods, is weakly ferromagnetic.    Furthermore, it is a unique type of itinerant ferromagnet in which the interlayer electrons not attached to any ion, rather than the 4$d$ electrons of Y, are responsible for the spin polarization.    

\begin{figure}
\includegraphics[width=8.5 cm]{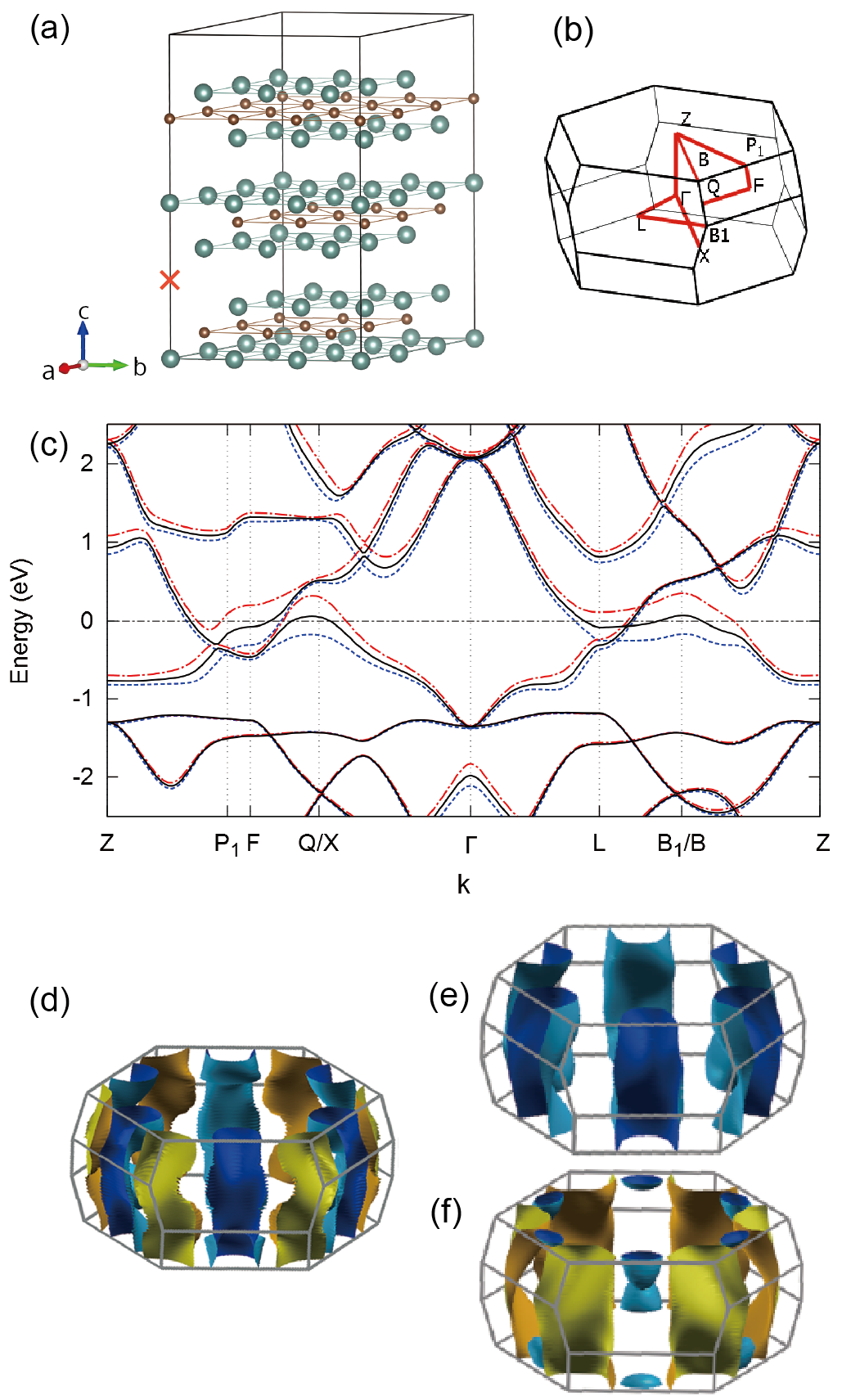}
\caption{\label{}(Color online)  (a) Crystal structure (anti-CdCl$_2$ structure) of Y$_2$C.  A supercell consisting of 3$\times$3$\times$1 conventional (hexagonal) unit cells is shown.  The large and small spheres represent Y and C atoms, respectively.   A representative interstitial site is indicated by a red cross.  A conventional unit cell (Y$_6$C$_3$) is made of three primitive (rhombohedral) cells (Y$_2$C).   (b) Brillouin zone of Y$_2$C.  B$_1$ is equivalent to B, and X is equivalent to Q by symmetry.  (c) Band structure obtained by spin-nonpolarized (black solid curves) and spin-polarized  (blue dashed curves and red dash-dotted curves) calculations using the PAW method.  The zero energy is taken to be the Fermi level.  (d) Fermi surfaces obtained from a spin-nonpolarized calculation (paramagnetic state).  (e) and (f) Fermi surfaces for the minority and majority spin channels, respectively, obtained from a spin-polarized calculation (ferromagnetic state).  The electron (hole) pockets are colored in blue (yellow). The darker side of the surfaces corresponds to lower energy.}
\end{figure}

Structure: Figure~1(a) illustrates the crystal structure of Y$_2$C. \cite{At2,Ma} A conventional (hexagonal) unit cell of Y$_2$C consists of nine  atomic layers having the same triangular 2D lattice structure, albeit being staggered.   The nine layers are grouped into three units, each made of three closely spaced layers (Y-C-Y) stacked with a layer separation of 1.35 \AA.    These units are separated by a distance (3.45~{\AA} between Y layers) larger than that between layers within the same unit (1.35~{\AA} between Y and C layers).   Since the standard oxidation numbers of Y and C are +3 and -4, respectively, these layer units are positively charged ([Y$_2$C]$^{2+}$) and the excess electrons are accumulated between the units to form a quasi-2D electride.\cite{In, Tad}  The whole crystal therefore has the [YCY]$^{2+}\cdot $2e$^- \cdot $[YCY]$^{2+} \cdot $2e$^- \cdot$ [YCY]$^{2+}$.... stacking structure.    As shown below, the interlayer electrons are loosely bound around the interstitial sites [indicated by a red cross in Fig.~1(a)] and extend in the layer plane to form a 2D electron system.  

Method of Calculation: We employed two density functional-based methods using  exchange-correlation functionals in the generalized gradient approximation (GGA)  : (1)  the full-potential linearized augmented plane wave (FLAPW) method,  as implemented in the All Electron Band Structure Calculation Package (ABCAP), \cite{Ham} and (2) the projector augmented wave (PAW) method with a plane wave basis, as implemented in the Vienna Ab Initio Simulation Package (VASP).\cite{Kr,Kr2,Bl,Kr3}  The FLAPW method takes into account all the electrons on an equal footing and is ``the most accurate method for electronic structure at the present time.''\cite{Mar}  The PAW method treats core electrons as frozen and is therefore more efficient at the cost of losing some accuracy.  In our calculations, the 2$s$ and 2$p$ orbitals of C and the 4$s$, 4$p$, 5$s$ and 4$d$ orbitals of Y were taken into account,  while lower-energy orbitals were frozen and ignored.  The specific GGA potentials used were PBE\cite{PBE} and PW91 \cite{PW91} for (1) and (2), respectively.   To reinforce these calculations, we also carried out plane-wave-based calculations using screened Heyd-Scuseria-Ernzerhof (HSE06) hybrid functional method,\cite{HSE, HSE06} as implemented in VASP, which goes beyond the density  functional theory by mixing PBE exchange and exact exchange.

For FLAPW calculations, we employed cutoff energies of 163 eV and 653 eV for the wave functions and charge/potential, respectively, and k-integration was carried out in the conventional (hexagonal) Brillouin zone (BZ) using a  $12\times 12 \times 4$ $\Gamma$-centered sampling mesh.  For PAW calculations, the plane waves were cut off at 800 eV, and integration over the primitive (rhombohedral) BZ was performed using a $17\times 17 \times 17$ $\Gamma$-centered mesh.  The hybrid calculations were performed with  cutoff energy of 400 eV and a sampling mesh of  $16\times 16 \times 16$. For both spin-polarized and spin-nonpolarized cases, the structure was relaxed using the PAW method until the force acting on each ion became less than 0.7 meV/\text{\AA}.  The optimized structures were used in FLAPW and hybrid functional calculations.   

\begin{table}
 \caption{\label{T1}Total energy decrease resulting from spin polarization $\Delta E$ and magnetization $m$, both per primitive unit cell (Y$_2$C), obtained by different calculation methods. Optimized lattice constants (spin-nonpolarized/spin-polarized) obtained by the PAW method are also shown.}
 \begin{ruledtabular}
 \begin{tabular}{lcccc}
 & $\Delta E$ & $m$  & $a$ & $c$ \\
 &(meV)&\textrm{ ($\mu_\text{B}$)} &(\AA)& (\AA)\\
\colrule
  PAW + PBE \footnotemark[1]& 15 & 0.38 & 3.61/3.62  & 18.45/18.42 \\ 
 FLAPW + PW91\footnotemark[2]& 12 & 0.36  & - & -  \\
 HSE06 \footnotemark[3]& 17 & 0.33  & - & -  \\
\end{tabular}
 \end{ruledtabular} 
  \footnotetext[1]{ Ref.~\onlinecite{PBE}}
 \footnotetext[2]{ Ref.~\onlinecite{PW91}}
\footnotetext[3]{ Refs.~\onlinecite{HSE, HSE06}}
 \end{table}

Results and Discussion: Table I compares the total energy decrease resulting from spin polarization $\Delta E$ and the magnetization $m$ per primitive unit cell (Y$_2$C)  obtained by the three methods.  Here,  $\Delta E$ is defined as $E_{tot}^{nonpol}-E_{tot}^{pol}$, where $E_{tot}^{pol}$ and $E_{tot}^{nonpol}$ denote the total energies calculated with and without spin polarization, respectively.     
As can be seen from the Table,  the results obtained by different methods are in very good agreement and predict Y$_2$C to be a weak itinerant ferromagnet.   The optimized lattice constants $a$ and $c$ (last two columns of Table I) are virtually unaffected by spin polarization. Hereafter, we will focus on the results obtained by the PAW method unless stated otherwise. 

Having two excess electrons per unit cell ([Y$_2$C]$^{2+}\cdot$ 2e$^-$]), Y$_2$C  would be a band insulator if there were no band overlap.  
The actual band structure calculated without spin polarization [black solid lines in Fig.~1(c)]  is semimetallic with two conduction bands overlapping near the Fermi level $E_F$ (= 0 throughout this paper).    This semimetallic band structure results in electron and hole pockets near the BZ edge.  As can be seen from the plots of Fermi surfaces in Fig.~1(d), the electron pockets include the P$_1$, F and L points, and the hole pockets include the Q$\equiv$X  and B$\equiv$B$_1$ points at the boundary of the BZ.  To identify the interlayer states, we placed an empty sphere of  1.8~{\AA}  radius at the interstitial site inside the interlayer gap [red cross in Fig.~1(a)] and projected the wave functions onto this sphere to obtain the partial density of states (PDOS) for the interstitial site.   The result shown in Fig.~2(c) indicates that the interlayer states span a rather narrow energy range of -1 to +0.4 eV.   Inspection of the charge density for these states reveals that it is indeed confined in the interlayer gap with a maximum of approximate $s$ symmetry at the interstitial site. Therefore, the interlayer band may be viewed as arising from the 2D lattice of overlapping virtual orbitals (quasiatoms) having $s$ symmetry at interstices.  Figure 2 also presents the (a) total density of states (DOS) and (b) PDOS for Y.   From the latter, one can see that the Y $d$ states are hybridized with the interlayer states near E$_F$.     An expanded plot (not shown) reveals that the overall bandwidth of the Y $d$ states is about 10 eV.  

\begin{figure}
\includegraphics[width=8.5 cm]{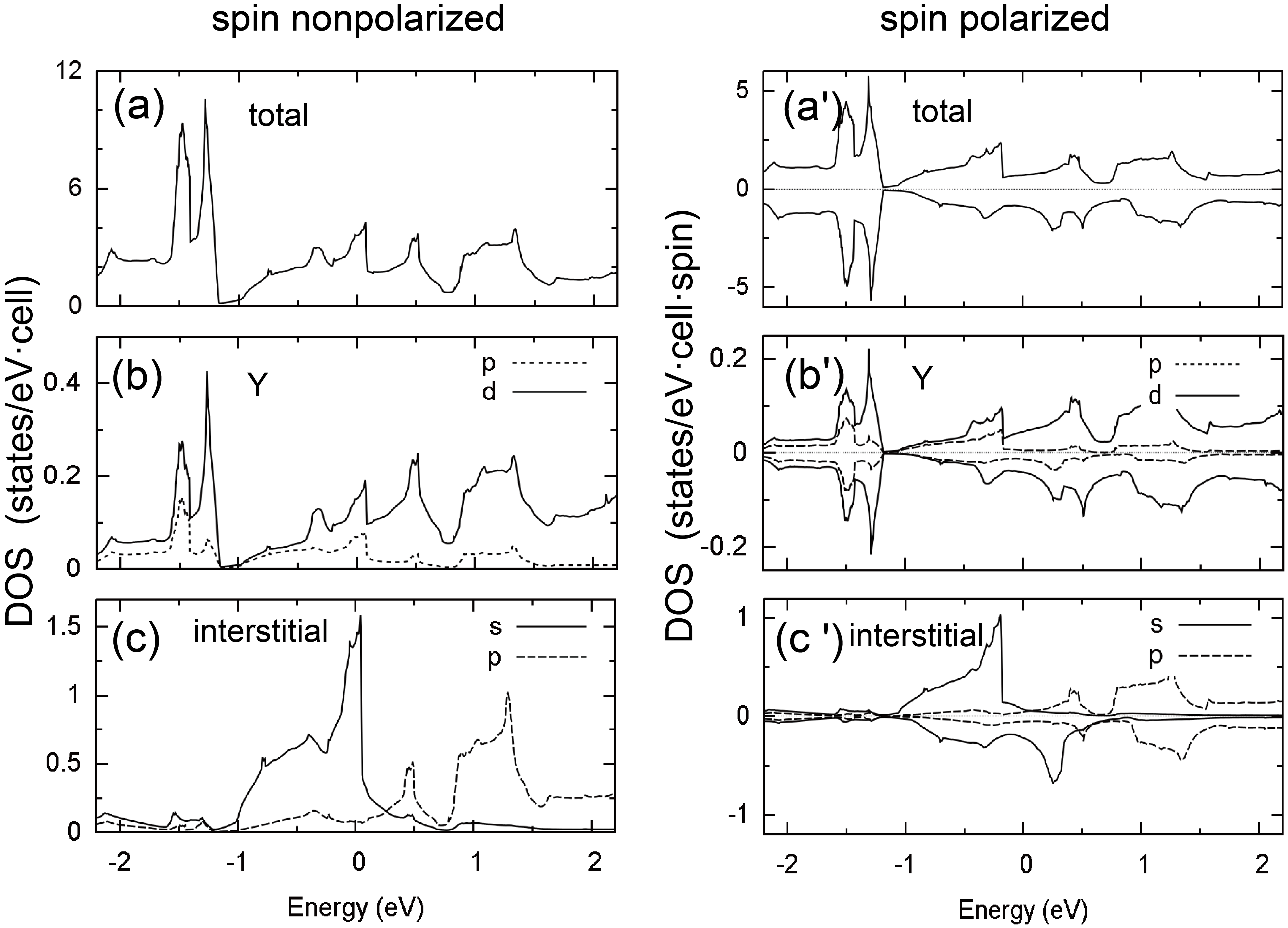}
\caption{\label{} (a) Total DOS, (b) PDOS for Y and (c) PDOS for the interstitial site calculated by spin-nonpolarized calculations using the PAW method.   Corresponding plots obtained from spin-polazired calculations are shown in (a'), (b') and (c'), respectively.  The positive and negative DOSs for the spin-polarized case denote the majority and minority spin channels, respectively. }
\end{figure}

The inclusion of spin polarization splits the interlayer bands, as shown by the blue dashed curves (majority spin channel)  and the red dash-dotted curves (minority spin channel) in Fig.~1(c),  causing the DOS peak to split and shift away from  $E_F$,  stabilizing the ferromagnetic state as can be seen in Figs.~2(a')-(c').   As the peak shifts, its shape undergoes a larger distortion than normally observed in conventional itinerant magnets such as iron.    In the ferromagnetic state, there is only one majority spin band crossing the Fermi level and this band has electron pockets enclosing P$_1$, F and L [Fig.~1(e)].  For the minority spin channel, the lower conduction band has hole pockets enclosing Q$\equiv$X and B$\equiv$B$_1$, and the upper conduction band has small electron pockets slightly off  P$_1$ towards Z [Fig.~1(f)].   

\begin{figure}
\includegraphics[width=8.5 cm]{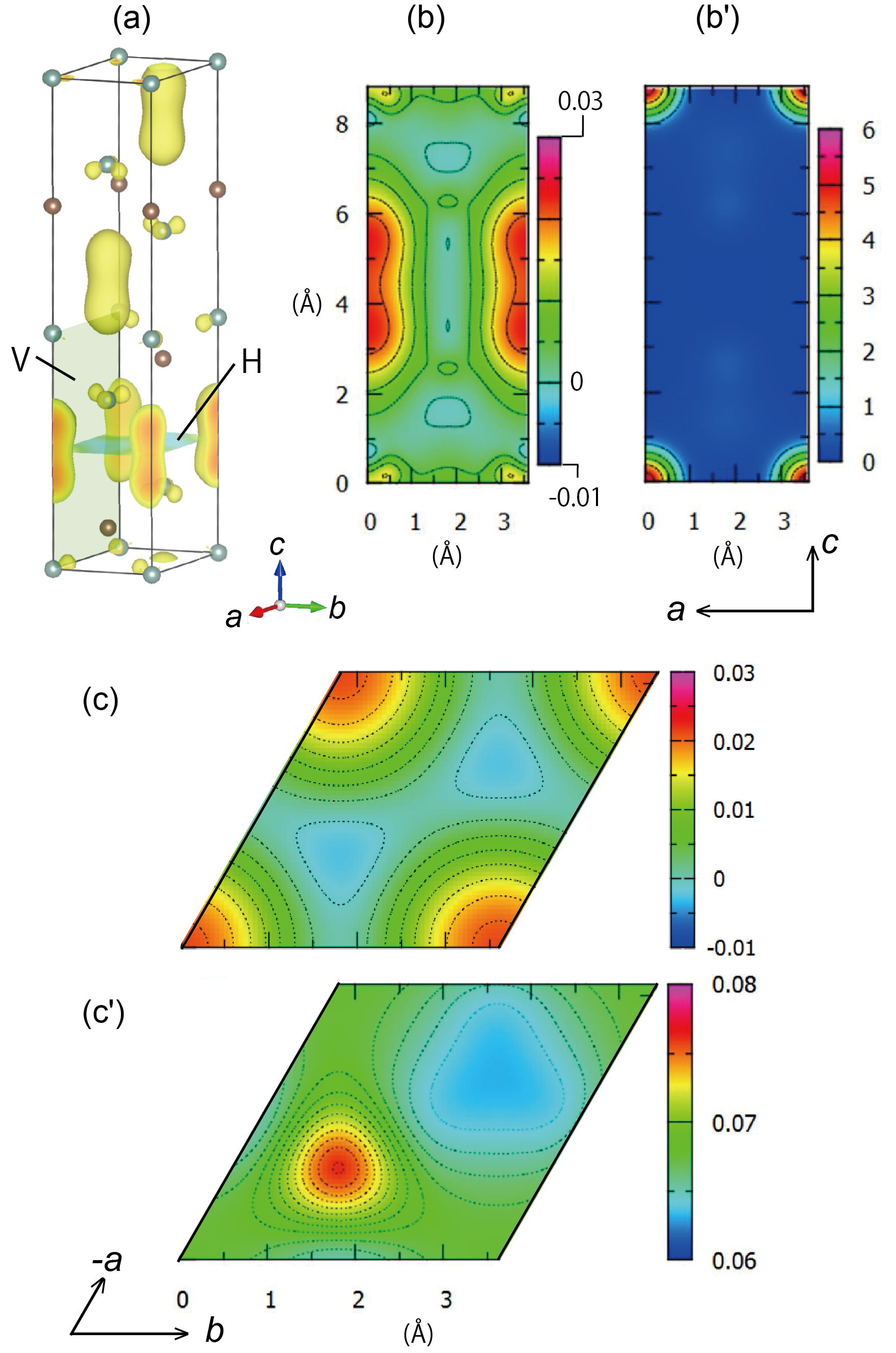}
\caption{\label{}(Color online) (a) Calculated magnetization density isosurface at 0.015 $\mu_B/$\AA$^3$.  The blue (brown) spheres represent Y (C) atoms.  (b) and (c) Magnetization densities in the (001) plane [indicated as H in (a)] and $a-c$ plane [indicated as V in (a)], respectively.  (b') and (c') represent corresponding plots for valence charge density.}
\end{figure} 

Figure~3(a) shows the magnetization density isosurface for the ferromagnetic state.    Remarkably, the bulk of the magnetization is concentrated around the interstitial sites, while very small polarization exists at Y sites.  (The Y ions are drawn with smaller spheres to make their polarization visible.)   This can be seen more clearly in the cross-sectional plots of magnetization in Figs.~3(b) and (c).  Here, (b) shows the cross section in the (001) horizontal plane  [denoted as H in Fig.~3(a)] passing through interstitial sites at its four corners, while (c) depicts the cross section in the $a-c$ vertical plane denoted as V in Fig.~3(a).  For comparison, corresponding plots of the charge density are shown in Figs.~3(b') and (c').    It can be seen that the magnetization is concentrated in off-ion regions where the charge density has its maxima.   The magnetic moment of Y ions (evaluated using the ionic radius for Y$^{3+}$ of 0.9 \AA) accounts for only 7.1~\% of the total magnetization.  This weak polarization of Y is likely to be the result of hybridization with the interlayer electrons.  These features indicate that the ferromagnetism of Y$_2$C is induced by the spin polarization of the electride electrons floating between the ionic layers.   

This picture of ferromagnetism induced by off-ion electrons is unconventional.  To ascertain that the $d$ electrons of Y are not playing the central role in the ferromagnetism, we also carried out calculations with on-site Coulomb interaction $U$ included in the Y $d$ orbitals.   It was found that magnetization decreases, rather than increases, as $U$ is increased from 0, which would be difficult to understand if the $d$ electrons are driving the ferromagnetism.   

In a wide range of itinerant ferromagnets, the ferromagnetism is understood to arise from a Stoner-type instability,\cite{St} i.e., energy is lowered by the exchange splitting of a sharp DOS peak around $E_F$.  The criterion for this instability is $D(E_F) I >1$ where $I$ is the exchange parameter.  This condition, derived originally using a simple molecular field argument postulating intra-atomic exchange, has been shown to be much more general and is applicable to both intra-atomic and more extended exchange interactions between the electrons.\cite{Kub} This Stoner-type mechanism appears to operate in Y$_2$C as can be seen from the DOS in Fig.~2.\cite{sh}   Recently, Pickard and Needs \cite{Pic} have made an important  prediction based on the density functional theory that the interstitial electrons in alkali metals at high pressures (also electrides) would exhibit a Stoner-type instability towards ferromagnetism.  Our results indicate that the interlayer electrons in Y$_2$C may show a similar instability at {\it ambient} pressure.  

\begin{figure}
\includegraphics[width=8.5 cm]{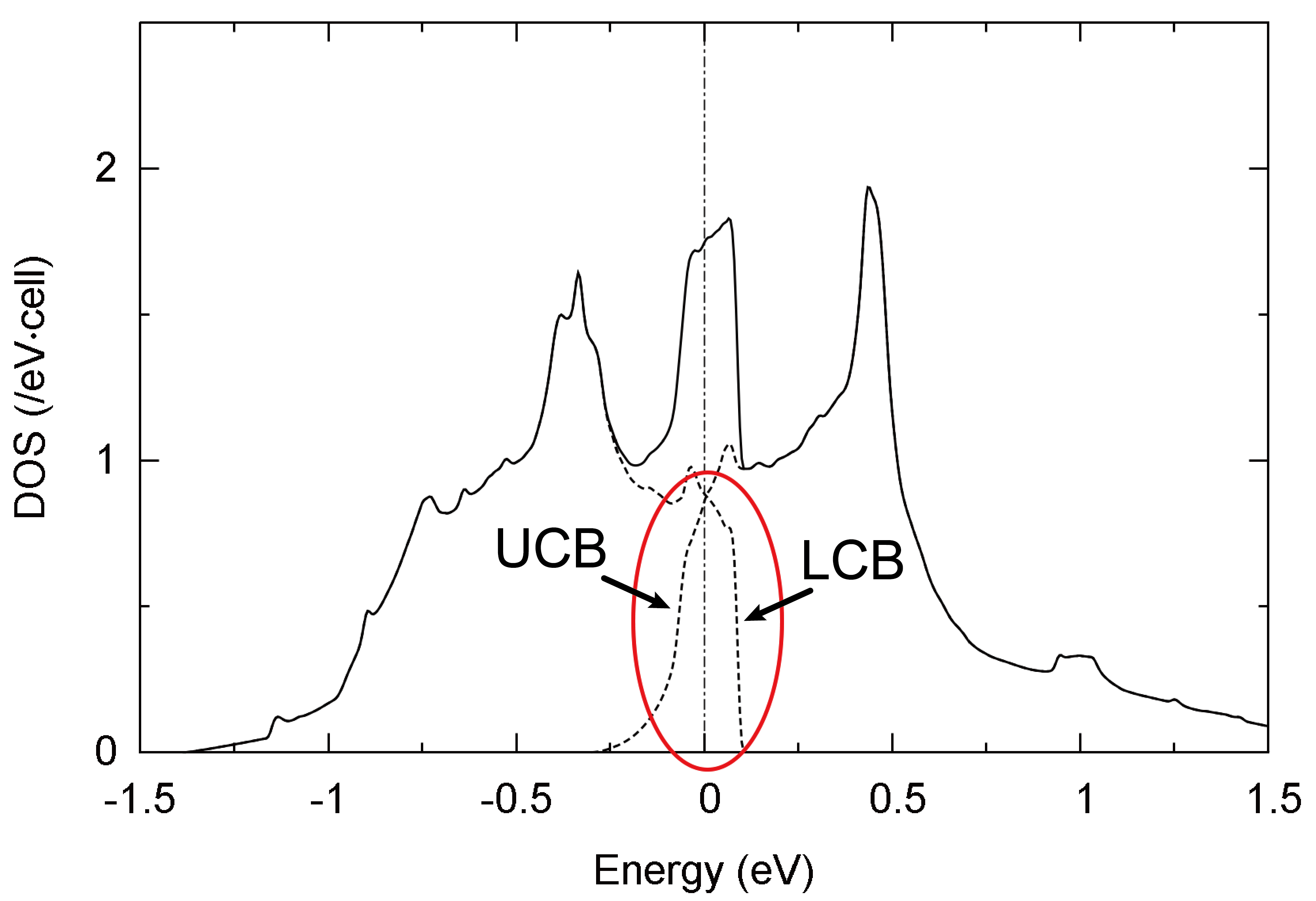}
\caption{\label{}(Color online) DOSs of the lower conduction band (LCB) and upper conduction band (UCB) obtained by a spin-nonpolarized FLAPW calculation (dashed curves).  As a result of band overlapping, highlighted by a red circle, the total DOS (solid curve) is enhanced near $E_F$.}
\end{figure}

One may wonder what causes the DOS to have a peak around $E_F$.   The dashed curves in Fig.~4 denote the DOSs of the lower conduction band (LCB) and upper conduction band (UCB) for the spin-nonpolarized state.   Both  curves drop sharply to zero in a step function manner toward the edge (highlighted by a red circle), which is characteristic of a 2D electron system.   As Y$_2$C is a semimetal, the two bands overlap near $E_F$ and cause their sum (total DOS, solid curve) to have a peak at $E_F$.   Such an enhancement of DOS at $E_F$ is expected to occur generally in semimetals with 2D conduction bands. 

In summary, ab initio calculations show the ground state of Y$_2$C to be weakly ferromagnetic, where the ferromagnetism is induced by electride electrons accumulated in the interlayer void space.    The magnetization measurement of polycrystalline Y$_2$C by Zhang \textit{et al.} \cite{Zh} shows  no ferromagnetic transition down to 2~K.    It is not yet clear whether there is a transition at an even lower temperature or there is no transition.  Even if the latter is the case, the system should be close to a ferromagnetic instability,  and spin fluctuations associated with this instability may be observable as an anomaly in spin susceptibility (deviation from the Curie-Weiss Law) and specific heat, \cite{Mor} or by techniques such as neutron diffraction and magnetic resonance.   The sharp rise in susceptibility below 20~K observed by Zhang {\it et al.}\cite{Zh} may signal such a fluctuation effect.

T. I.  thanks Drs. Norio Ota, Susumu Saito and Xiao Zhang for useful discussions. This work was supported by the MEXT Elements Strategy Initiative to Form Core Research Center and the JST ACCEL Program.  The Fermi surfaces were plotted by XCrySDen.\cite{XCr}


\begin{thebibliography}{}
\bibitem{sh} This is consistent with our finding that the ferromagnetism of Y$_2$C is suppressed by shifting $E_F$ away from the DOS peak in both positive and negative directions.
\end{thebibliography}
\end{document}